\documentclass[11pt]{article}
\usepackage{graphicx}
\usepackage{float} 
\usepackage{amsmath}
\usepackage{amsfonts}   
\usepackage{amssymb}

\begin{document}
\date{}
\title{{\bf{\Large Second Order Phase Transition and Thermodynamic Geometry in Kerr-AdS Black Hole}}}
\author{
 {\bf {\normalsize Rabin Banerjee}$
$\thanks{e-mail: rabin@bose.res.in}},\, 
 {\bf {\normalsize Sujoy Kumar Modak}$
$\thanks{e-mail: sujoy@bose.res.in}}\\
 {\normalsize S.~N.~Bose National Centre for Basic Sciences,}
\\{\normalsize JD Block, Sector III, Salt Lake, Kolkata-700098, India}
\\[0.3cm]
{\bf {\normalsize Saurav Samanta}$
$\thanks{e-mail: srvsmnt@gmail.com}}\\
 {\normalsize Narasinha Dutt College}
\\{\normalsize 129, Belilious Road, Howrah-711101, India}
\\[0.3cm]
}

\maketitle


\begin{abstract}
We discuss a scheme based on Ehrenfest like equations to exhibit and classify transitions between two phases (with ``smaller'' and ``larger'' masses) of Kerr AdS black holes. We show that for fixed angular velocity this phase transition is of second order as both Ehrenfest's equations are satisfied. Finally we make a close connection of the results found from this analysis with those obtained from the thermodynamic state space geometry approach.  
\end{abstract}

\section{Introduction}
The pioneering works of Bekenstein and Hawking have opened many interesting aspects of unification of quantum mechanics, gravity and thermodynamics. These are known for the last forty years \cite{Beken1}-\cite{Hawking2}. Since then various thermodynamical properties of black holes have been studied widely and now we have a considerable understanding about the microscopic origin of these properties due to a pioneering work by Strominger and Vafa \cite{vafa}. 
Recently there has been a suggestion that fluctuation theory, whose origin is contained in statistical mechanics, can be described solely by thermodynamics \cite{Ruppeiner:1995zz}. This approach is based on the thermodynamic state space geometry \cite{Ruppeiner:1983zz,Ruppeiner:1995zz} which is commonly known as the Ruppeiner geometry.  For second order phase transitions Ruppeiner curvature scalar ($R$) is expected to diverge at the critical point \cite{Aman:2003ug,Ruppeiner:2008kd,Shen:2005nu,Sahay:2010wi}. Motivated by the success of this geometry to identify a phase transition in normal thermodynamic system, there have been several recent works \cite{Cai:1998ep}-\cite{Aman:2010me} which address the black hole phase transitions through this approach.

In this paper we discuss a new formulation to analyse phase transition in Kerr AdS black holes. It is based on an application of Clapeyron's and Ehrenfest's ideas to black hole systems. In our analysis we keep the angular velocity $\Omega$ (analog of pressure) to be constant which is essential for implementing the above ideas in Kerr-AdS black holes. Specifically, in Kerr AdS black hole (for fixed $\Omega$) we do not find any discontinuity in entropy and therefore the chance of a first order phase transition is automatically ruled out. In our analysis, we find a discontinuity in the specific heat at constant angular velocity ($C_{\Omega}$). Conventional works \cite{BHT}-\cite{Doneva:2010iq} attempt to identify a phase transition from this discontinuity but only at a qualitative level. One of the motivations of this paper is to provide a detailed quantitative analysis of this aspect. Before that, however, we highlight an  important point {\it i.e.} infinite divergence of all relevant quantities (including the specific heat) at the critical point. An infinite divergence of heat capacity was also discussed earlier in \cite{Cald}, for the Kerr-Newman-AdS black holes in a canonical ensemble. Note that in a standard second order phase transition these discontinuities are finite.  Although a discontinuity in the specific heat is necessary, it is not sufficient to correctly identify a phase transition. Indeed, it is essential to verify the Ehrenfest equations. These equations, in the context of black holes, were earlier derived in \cite{BMS}.  Because of the infinite divergence of all relevant quantities (including the specific heat) we develop a technique to study Ehrenfest's equations infinitesimally close to the critical point. It is found that both Ehrenfest's equations are satisfied. Consequently this phase transition is a genuine second order one. 

The remainder of our work concerns the application of the thermodynamic geometry approach \cite{Ruppeiner:1983zz,Ruppeiner:1995zz} to identify the occurrence of black hole phase transitions. We explicitly calculate and plot the Ruppeiner curvature scalar ($R$). The plots show the divergence of $R$ at the same critical point where specific heat capacity ($C_{\Omega}$) was diverging (corresponding to $\Omega < 1$). This behaviour is remarkable in the sense that these divergences are usually associated with second order phase transitions in standard thermodynamical systems. Indeed our study proves that the divergence found here indicates a genuine second order phase transition in Kerr-AdS black holes. Furthermore we find some cases (when $\Omega\geq 1$) where $R$ diverges but specific heat does not. This is also a striking observation where the divergence of $R$ does not imply any discontinuity in the specific heat. However in this case ($\Omega\geq 1$) the divergence of $R$ exhibits a special property. The left hand limit and the right hand limit of $R$ do not agree at the singular point. This behaviour is completely different from the previous case ($\Omega <1$) where both limits are same at the singular point (which is correctly associated with a phase transition point). This illustrates the crucial role of the nature of divergence of $R$ in identifying a discontinuity in the specific heat. Our analysis therefore reveals important implications of thermodynamic geometry in identifying and classifying phase transitions in black holes as compared to usual thermodynamic systems.



\section{The Kerr AdS black hole}
Kerr AdS black hole is a solution of Einstein equation in (3+1) dimensions with a negative cosmological constant $\Lambda=-\frac{3}{l^2}$. It is characterised by two parameters, namely mass $M$ and angular momentum $J$. The first law of thermodynamics for the Kerr-AdS black hole is given by \cite{gib}
\begin{equation}
dM=TdS+\Omega dJ
\label{fla}
\end{equation}
where $T$ is the Hawking temperature and $\Omega$ is the difference between the angular velocities at the event horizon ($\Omega_H$) and at infinity ($\Omega_{\infty}$) \cite{Cald}-\cite{Cardoso}. The explicit expressions for various parameters are given in \cite{Cald}-\cite{Cardoso}. We write these equations with a suitable rescaling where we define $Tl,\frac{M}{l},\frac{J}{l^2}, \frac{S}{l^2}$, $\Omega l$ as the temperature ($T$), Mass ($M$), angular momentum ($J$), entropy ($S$) and angular velocity ($\Omega$) of the black hole. This is done so that $l$ no longer appears in any equation. In terms of these newly defined variables we find,
\begin{eqnarray}
T = \frac{1}{8\pi M}\left(1-\frac{4\pi^2J^2}{S^2}+\frac{4S}{\pi}+\frac{3S^2}{\pi^2}\right),
\label{temp}
\end{eqnarray}
\begin{eqnarray}
M^2&=&\frac{S}{4\pi}+\frac{\pi J^2}{S} + J^2+\frac{S}{2\pi}\left(\frac{S}{\pi}+\frac{S^2}{2\pi^2}\right)\label{M2}\\
\Omega&=&\frac{\pi J}{MS}\left(1+\frac{S}{\pi}\right)\\
\frac{J}{S}&=&\frac{M\Omega}{\pi+S}\label{JS}\\
a&=&\frac{J}{M}=\frac{S\Omega}{\pi}\left(1+\frac{S}{\pi}\right)^{-1}\label{12}
\end{eqnarray}
Now using (\ref{JS}) we substitute $J$ in (\ref{M2}) to express $M^2$ in terms of $S$ and $\Omega$. This yields
\begin{eqnarray}
M^2 =\frac{S}{4\pi}\frac{1+\frac{2S}{\pi}(1+\frac{S}{2\pi})}{1-\frac{\Omega^2 S}{\pi(1+\frac{S}{\pi})}}.
\label{Msquare}
\end{eqnarray}
Likewise, replacing $M$ (\ref{Msquare}) in (\ref{temp}), the semiclassical temperature is found to be,
\begin{eqnarray}
T=\sqrt{\frac{S(\pi+S)^3}{(\pi+S-S\Omega^2)}}\left[\frac{\pi^2-2\pi S(\Omega^2-2)-3S^2(\Omega^2-1)}{4\pi^{\frac{3}{2}}S(\pi+S)^2}\right].
\label{14}
\end{eqnarray}
From the above equation we see that $T$ is real only when
\begin{eqnarray}
\pi+S-S\Omega^2>0
\label{cond}
\end{eqnarray}
which implies $\Omega^2<1+\frac{\pi}{S}$. This imposes a restriction on $\Omega$ for a fixed value of the entropy. To get further insight into the thermodynamical behaviour of the black hole we calculate the semiclassical specific heat at constant angular velocity (which is the analogue of $C_{P}$) from (\ref{14}) by using the relation $C_{\Omega}=T\left(\frac{\partial S}{\partial T}\right)_{\Omega}=\frac{T}{\left(\frac{\partial T}{\partial S}\right)_{\Omega}}$. This is found to be 
\begin{eqnarray}
C_{\Omega}=\frac{2S(\pi+S)(\pi+S-S\Omega^2)\left(\pi^2-2\pi S(\Omega^2-2)-3S^2(\Omega^2-1)\right)}{(\pi+S)^3(3S-\pi)-6S^2(\pi+S)^2\Omega^2+S^3(4\pi+3S)\Omega^4}
\label{15}
\end{eqnarray}
With these expressions for the temperature (\ref{14}) and specific heat capacity (\ref{15}) we now proceed to the next section where the phase transition phenomena will be discussed.


\section{Ehrenfest's scheme and phase transition}
In this section we shall use (and improve) the approach developed in \cite{BMS} to study phase transition in a black hole. This approach, based on standard thermodynamical tools, deals with the identification and classification of a phase transition in a black hole.   

Let us first plot (\ref{14}) and (\ref{15}) with respect to $S$ for a fixed value of $\Omega$ ($\Omega=0.3$). These plots are depicted in figure 1. From figure 1(a) we find that the semiclassical Hawking temperature ($T$) is continuous when plotted with the semiclassical entropy ($S$). For a first order phase transition (like liquid to vapour) the first order derivative of Gibbs free energy, i.e. volume and entropy are discontinuous. The continuity of entropy ($S$) (and also $J$) suggests that for fixed $\Omega$ there is no first order phase transition taking place in the Kerr AdS black hole. Nevertheless there is a minimum temperature for a certain value of $S$ ($S=1.09761$). Figure 1(b) shows a discontinuity in $C_{\Omega}$ at this critical value of entropy ($S=S_c=1.09761$). However in this plot the discontinuity of specific heat is infinite which is completely different from the ordinary thermodynamical systems  (for example feromagnetic to paramagnetic transformation) where finite discontinuity occurs. Note that this transition between two phases of Kerr-AdS black holes have different entropies. As entropy is proportional to the square of the black hole mass the critical point $S_{c}$ is essentially seperating two branches of Kerr-AdS black holes with ``smaller'' and ``larger'' masses. In an earlier work \cite{Cald} an infinite divergence of specific heat was reported in the case of Kerr-Newman-AdS black hole in a canonical ensemble. The possibility of a second order phase transition in this case was also stated.

\begin{figure}[ht]
\centering
\includegraphics[angle=0,width=8cm,keepaspectratio]{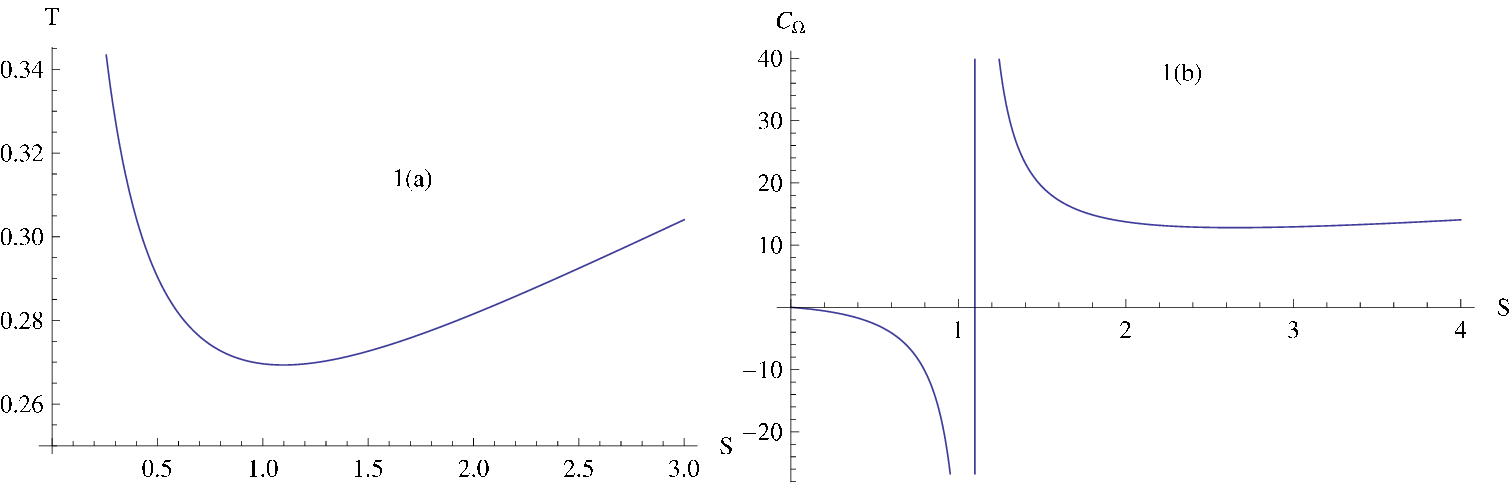}
\caption[]{{Semi-classical Hawking temperature ($T$) and Specific heat ($C_{\Omega}$) vs entropy ($S$) plot for fixed $\Omega=0.3$.}}
\label{figure 1}
\end{figure}


The conventional way to  understand a phase transition phenomena where the entropy is continuous and the specific heat capacity is discontinuous is based on the standard Ehrenfest's prescription. Note that for any true second order phase transition both Ehrenfest's equations must be satisfied at the critical point. Because of the differences between conventional systems and black holes, care must be exercised in applying Ehrenfest's scheme to study phase transitions in black holes. In the remaining part of this section we shall follow this spirit to uncover the properties of the phase transition depicted in figures 1(a) and 1(b).

Let us now recall Ehrenfest's scheme of analysing phase transition which is done in usual thermodynamics. For a black hole system the two Ehrenfest's equations were derived in our earlier work \cite{BMS}. Ehrenfest's first and second equations, for the Kerr AdS black hole, are given by \cite{BMS} 
\begin{eqnarray}
-\left(\frac{\partial\Omega}{\partial T}\right)_S=\frac{C_{\Omega_2}-C_{\Omega_1}}{TJ(\alpha_2-\alpha_1)}
\label{16}
\end{eqnarray}
and
\begin{eqnarray}
-\left(\frac{\partial\Omega}{\partial T}\right)_J=\frac{(\alpha_2-\alpha_1)}{(k_{T_2}-k_{T_1})}
\label{17}
\end{eqnarray}
where,
\begin{equation}
\alpha=\frac{1}{J}\left(\frac{\partial J}{\partial T}\right)_{\Omega}, 
\label{18}
\end{equation}
\begin{equation}
k_{T}=\frac{1}{J}\left(\frac{\partial J}{\partial {\Omega}}\right)_T.
\label{19}
\end{equation}
The above quantities (\ref{18}) and (\ref{19}) were also defined in \cite{monteiro,monteiro2} to discuss the thermodynamic stability of black holes.

We noted earlier that there is a discontinuity in $C_{\Omega}$ as shown in figure 1(b). Nevertheless, presence of discontinuities in $\alpha$ and $k_T$ are also necessary to show that a phase transition is taking place. Since $C_{\Omega}$, $\alpha$ and $k_T$ are all second order derivatives of the Gibbs free energy, they must be discontinuous in a second order phase transition. So in order to proceed further we calculate these physical entities and plot them. First using (\ref{JS}) and (\ref{Msquare}) we can express $J$ as a function of $S$ and $\Omega$, as given by
\begin{eqnarray}
J(S,\Omega)=\frac{S\Omega\sqrt{\frac{S(\pi+S)^3}{(\pi+S-S\Omega^2)}}}{2\pi^{\frac{3}{2}}(\pi+S)}.
\label{26}
\end{eqnarray}
Using the definition of $a$ (\ref{12}) we write (\ref{18}) as,
\begin{eqnarray}
J\alpha=M \left(\frac{\partial a}{\partial T}\right)_{\Omega}+a\left(\frac{\partial M}{\partial T}\right)_{\Omega}
\label{20}
\end{eqnarray}
Now to calculate the first term of the right hand side we need a functional relationship between $a, T$ and $\Omega$. To do so we rewrite the semiclassical entropy of the Kerr AdS black hole (\ref{JS}) in terms of $\Omega$ and $a$ as
\begin{eqnarray}
S=\frac{\pi a}{\Omega-a}.
\label{21}
\end{eqnarray}
In (\ref{14}) $T$ was expressed as $T=T(S,\Omega)$. Substituting (\ref{21}) in (\ref{14}) we write the temperature in terms of $a$ and $\Omega$ as,
\begin{eqnarray}
T=T(\Omega,a)=\frac{2a\Omega^2+\Omega a^2-2a-\Omega}{4\pi\sqrt{a(a-\Omega)(a\Omega-1)}}.
\label{22}
\end{eqnarray}
Now it is straightforward to calculate $\left(\frac{\partial a}{\partial T}\right)_{\Omega}$. To calculate the second term of the right hand side of (\ref{20}), we write $\left(\frac{\partial M}{\partial T}\right)_{\Omega}=\left(\frac{\partial M}{\partial S}\right)_{\Omega}\left(\frac{\partial S}{\partial T}\right)_{\Omega}$. Now using (\ref{Msquare}) and (\ref{14}) we find, $\left(\frac{\partial M}{\partial T}\right)_{\Omega}$. Making use of these results, the value of $a$ from (\ref{21}) in (\ref{20}) and substituting $J$ from (\ref{26}) we finally obtain
the analog of volume expansion coefficient,
\begin{eqnarray}
\alpha=\frac{\sqrt{4\pi^3 S(\pi+S)(\pi +S-S\Omega^2)}[6\Omega(\pi+S)^2-2S\Omega^2(2\pi+3S)]}{(\pi+S)^3(3S-\pi)-6S^2(\pi+S)^2\Omega^2+S^3(4\pi+3S)\Omega^4}
\label{25alp}
\end{eqnarray}
  
Now to find an expression for the analog of compressibility $k_T=\frac{1}{J}\left(\frac{\partial J}{\partial {\Omega}}\right)_T$, we first write $J$ in terms of $a$ and $\Omega$ in the following manner,
\begin{eqnarray}
J=\frac{1}{2}\sqrt{\frac{a(\Omega-a)}{(1-a\Omega)}}\left[\frac{a\Omega}{(\Omega-a)^2}\right].
\label{26aa}
\end{eqnarray}
Since $J$ cannot be expressed in terms of $T$ and $\Omega$ we shall use the rules of partial differentiation to find $k_T$ from (\ref{22}) and (\ref{26aa}). From the theorem $dJ=\left(\frac{\partial J}{\partial a}\right)_{\Omega}da+\left(\frac{\partial J}{\partial \Omega}\right)_{a}d\Omega$ we can write
\begin{eqnarray}
\left(\frac{\partial J}{\partial\Omega}\right)_{T}=\left(\frac{\partial J}{\partial a}\right)_{\Omega}\left(\frac{\partial a}{\partial \Omega}\right)_T+\left(\frac{\partial J}{\partial \Omega}\right)_{a}
\label{301}
\end{eqnarray}
The above equation is written in a more useful form by substituting $\left(\frac{\partial a}{\partial \Omega}\right)_T=-\frac{\left(\frac{\partial T}{\partial \Omega}\right)_a}{\left(\frac{\partial T}{\partial a}\right)_{\Omega}}$. This gives
\begin{eqnarray}
\left(\frac{\partial J}{\partial\Omega}\right)_{T}=\frac{\left(\frac{\partial J}{\partial \Omega}\right)_{a}\left(\frac{\partial T}{\partial a}\right)_{\Omega}-\left(\frac{\partial T}{\partial \Omega}\right)_{a}\left(\frac{\partial J}{\partial a}\right)_{\Omega}}{\left(\frac{\partial T}{\partial a}\right)_{\Omega}}.
\end{eqnarray}
Using (\ref{22}) and (\ref{26aa}) in the above equation we obtain $k_T$ in terms of $a$ and $\Omega$. Finally eliminating $a$ in favour of $S$ and $\Omega$ by using (\ref{21}) and then substituting $J$ from (\ref{26}), we find,
\begin{eqnarray}
k_T &=& \frac{(3S-\pi)(\pi+S)^3+2S(\pi+S)^2(4\pi+3S)\Omega^2-S^2(2\pi+3S)^2\Omega^4}{(3S-\pi)\Omega(\pi+S)^3-6S^2(\pi+S)^2\Omega^3+S^3(4\pi+3S)\Omega^5}\nonumber\\
\label{JKT}
\end{eqnarray}

\begin{figure}[ht]
\centering
\includegraphics[angle=0,width=8cm,keepaspectratio]{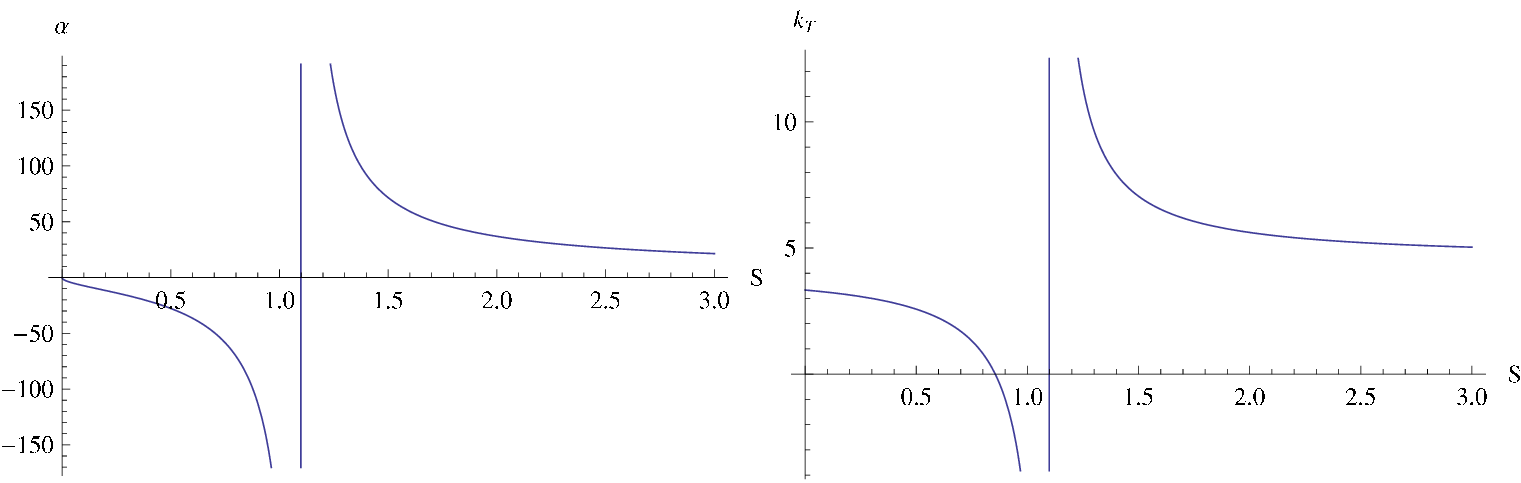}
\caption[]{{$\alpha$ and $k_T$ vs entropy ($S$) plot for fixed $\Omega=0.3$.}}
\label{figure 2}
\end{figure}

Let us now plot (\ref{25alp}) and (\ref{JKT}) with respect to $S$ for a fixed value of $\Omega=0.3$. These two plots are shown in figure 2 and they show a discontinuity in both these quantities at the same critical value of $S_c=1.09761$ (see figure 1(a)). With these results we are now convinced about a genuine phase transition in Kerr-AdS black hole.


We shall now perform a numerical analysis to understand the order of this phase transition. The idea is to investigate the two Ehrenfest's equations and check their validity. For any true second order phase transition both Ehrenfest's equations must be satisfied at the critical point.  However, because of the infinite divergences of all relevant quantities at that point one should be very careful in doing the numerical computations.  

We start by considering the first Ehrenfest's equation (\ref{16}). The left hand side of (\ref{16}) can be calculated easily from the relation (\ref{14}). This is found to be 
\begin{eqnarray}
-\left(\frac{\partial\Omega}{\partial T}\right)_S=-\frac{4\pi^{\frac{3}{2}}(\pi+S-S\Omega^2)^2\sqrt\frac{S(\pi+S)^3}{(\pi+S-S\Omega^2)}}{S\Omega\left(S(\pi+S)(2\pi+3S)\Omega^2-3(\pi+S)^3\right)}.
\label{31}
\end{eqnarray}
At the phase transition point we have $S=1.09761$, $\Omega=0.3$. Using these values in (\ref{31}) we obtain,
\begin{eqnarray}
-\left(\frac{\partial\Omega}{\partial T}\right)_S=23.2085
\label{31a}
\end{eqnarray}
which yields the left hand side of the first Ehrenfest's equation (\ref{16}). 

To calculate the right hand side recall the expressions of $C_{\Omega}$ and $\alpha$ from (\ref{15}) and (\ref{25alp}). They have the form $C_{\Omega}=\frac{f(S,\Omega)}{g(S,\Omega)}$ and $\alpha=\frac{h(S,\Omega)}{g(S,\Omega)}$ where the denominators of both quantities are identical, given by $g(S,\Omega)=(\pi+S)^3(3S-\pi)-6S^2\Omega^2(\pi+S)^2+S^3\Omega^4(4\pi+3S)$. Note that for a fixed value of angular velocity (say $\Omega=0.3$) both $C_{\Omega}$ and $\alpha$ diverge when $g(S_c)=0$. Since this is polynomial equation with maximum power 4 there are four roots of this equation. Among these two are real and other two are imaginary. The real roots are positive and negative. Since entropy by definition is always positive, $S_c=1.09761$ is the only root which is contributing physically. This is the phase transition point where both Ehrenfest's relations are applicable. Let us now expand the denominators of $C_{\Omega}$ and $\alpha$ infinitesimally close to the critical point ($S_c$) in the following manner
\begin{eqnarray}
C_{\Omega} &=& \frac{f(S_c)}{g'(S_c)[(S-S_c)+\frac{1}{2}(S-S_c)^2\frac{g''(S_c)}{g'(S_c)}+\cdot\cdot\cdot]}\label{expsp}\\
&&=\frac{4.82247}{(S-S_c)+\frac{1}{2}(S-S_c)^2\frac{g''(S_c)}{g'(S_c)}+\cdot\cdot\cdot},\label{exspnm}
\end{eqnarray}  
\begin{eqnarray}
\alpha &=& \frac{h(S_c)}{g'(S_c)[(S-S_c)+\frac{1}{2}(S-S_c)^2\frac{g''(S_c)}{g'(S_c)}+\cdot\cdot\cdot]}\label{expve}\\
&&=\frac{24.6136}{(S-S_c)+\frac{1}{2}(S-S_c)^2\frac{g''(S_c)}{g'(S_c)}+\cdot\cdot\cdot}\label{exvenm}.
\end{eqnarray}
Now consider two points $S_1=S_c+\epsilon$ and $S_2=S_c-\epsilon$ ($\epsilon<<1$) which are infinitesimally close to $S_c$ and are respectively connected with phase-1 and phase-2 of the curves given in figures 1(b) and (2). The values of $C_{\Omega}$ and $\alpha$ for corresponding phases are given by $C_{\Omega}|_{S_1}=C_{\Omega_1},~C_{\Omega}|_{S_2}=C_{\Omega_2}$ and $\alpha|_{S_1}=\alpha_1,~\alpha|_{S_2}=\alpha_2$ respectively. Consequently the right hand side of the first Ehrenfest's relation (\ref{16}) reads as
\begin{eqnarray}
\frac{C_{\Omega_2}-C_{\Omega_1}}{TJ(\alpha_2-\alpha_1)}=\frac{4.82247[\frac{1}{(S_2-S_c)+\frac{1}{2}(S_2-S_c)^2\frac{g''(S_c)}{g'(S_c)}}+\cdot\cdot\cdot-\frac{1}{(S_1-S_c)+\frac{1}{2}(S_1-S_c)^2\frac{g''(S_c)}{g'(S_c)}}+\cdot\cdot\cdot]}{TJ\times24.6136[\frac{1}{(S_2-S_c)+\frac{1}{2}(S_2-S_c)^2\frac{g''(S_c)}{g'(S_c)}}+\cdot\cdot\cdot-\frac{1}{(S_1-S_c)+\frac{1}{2}(S_1-S_c)^2\frac{g''(S_c)}{g'(S_c)}}+\cdot\cdot\cdot]}\label{rhfe}
\end{eqnarray}
Remarkably from the above equation we find that the divergences in $C_{\Omega}$ and $\alpha$ (inside the third brackets) near the critical point exactly cancel out for all order expansions and we are left with a finite result. Using the critical values $T(S_c)$ (found from (\ref{14})) and $J(S_c)$ (found from (\ref{26})), as one can check, the right hand side of the first Ehrenfest's relation is found to be
\begin{eqnarray}
\frac{C_{\Omega_2}-C_{\Omega_1}}{TJ(\alpha_2-\alpha_1)}=23.2085.
\label{rhval}
\end{eqnarray} 
A comparison between the above equation and (\ref{31a}) proves that the first Ehrenfest's relation is satisfied for the phase transition in Kerr-AdS black hole.

Let us now examine the validity of the second Ehrenfest's equation (\ref{17}). To calculate $\left(\frac{\partial\Omega}{\partial T}\right)_J$, equation (\ref{22}) is written as
\begin{eqnarray}
dT=\left(\frac{\partial T}{\partial a}\right)_{\Omega}da+\left(\frac{\partial T}{\partial \Omega}\right)_{a}d\Omega~.
\label{dt}
\end{eqnarray} 
For a process where $J$ is constant, from (\ref{26aa}), one has  
\begin{eqnarray}
\left(\frac{\partial J}{\partial a}\right)_{\Omega}da+\left(\frac{\partial J}{\partial \Omega}\right)_{a}d\Omega=0~.
\label{j0}
\end{eqnarray}
Now eliminating terms involving $da$ from the above two relations and then rearranging terms, yields,
\begin{eqnarray}
-\left(\frac{\partial T}{\partial\Omega}\right)_{J}=\frac{\left(\frac{\partial T}{\partial a}\right)_{\Omega}\left(\frac{\partial J}{\partial \Omega}\right)_{a}-\left(\frac{\partial T}{\partial \Omega}\right)_{a}\left(\frac{\partial J}{\partial a}\right)_{\Omega}}{\left(\frac{\partial J}{\partial a}\right)_{\Omega}}.
\label{29}
\end{eqnarray}
Finally using (\ref{22}) and (\ref{26aa}) and substituting $a$ in terms of $\Omega$ and $S$ from (\ref{21}) we find, 
\begin{eqnarray}
-\left(\frac{\partial\Omega}{\partial T}\right)_J&=&\sqrt{\frac{\pi+S-S\Omega^2}{S(\pi+S)^3}}\left[\frac{4\pi^{3/2}S(\pi+S)^2\Omega(\left(3(\pi+S)^2-S(2\pi+3S)\Omega^2\right))}{(\pi+S)^3(3S-\pi)+2S(\pi+S)^2(4\pi+3S)\Omega^2-S^2(2\pi+3S)^2\Omega^4}\nonumber\right]\\\label{32}
\end{eqnarray}
At the critical point ($S=1.09761$) and for the chosen value of $\Omega=0.3$, we obtain,
\begin{eqnarray}
-\left(\frac{\partial\Omega}{\partial T}\right)_J=23.2085,
\label{eren2l}
\end{eqnarray}
which gives the left hand side of the second Ehrenfest's equation (\ref{17}). 

In order to calculate the right hand side first let us write (\ref{JKT}) in the following manner $k_T=\frac{u(S,\Omega)}{g(S,\Omega)}$. For $\Omega=0.3$ expanding $k_T$ near the critical point ($S_c$) we find 
\begin{eqnarray}
k_T=\frac{1.06055}{(S-S_c)+\frac{1}{2}(S-S_c)^2\frac{g''(S_c)}{g'(S_c)}+\cdot\cdot\cdot}.
\label{expkt}
\end{eqnarray} 
Now repeating the steps as carried out for the right hand side of (\ref{16}) we finally obtain the right hand side of the second Ehrenfest's relation (\ref{17}) as,
\begin{eqnarray}
\frac{\alpha_2-\alpha_1}{k_{T_2}-k_{T_1}} &=& \frac{24.6136[\frac{1}{(S_2-S_c)+\frac{1}{2}(S_2-S_c)^2\frac{g''(S_c)}{g'(S_c)}}+\cdot\cdot\cdot-\frac{1}{(S_1-S_c)+\frac{1}{2}(S_1-S_c)^2\frac{g''(S_c)}{g'(S_c)}}+\cdot\cdot\cdot]}{1.06055[\frac{1}{(S_2-S_c)+\frac{1}{2}(S_2-S_c)^2\frac{g''(S_c)}{g'(S_c)}}+\cdot\cdot\cdot-\frac{1}{(S_1-S_c)+\frac{1}{2}(S_1-S_c)^2\frac{g''(S_c)}{g'(S_c)}}+\cdot\cdot\cdot]}\label{secrh}\\
&&=23.2083,\label{secrh1}
\end{eqnarray}
Thus from the above equation once again we find that the divergent terms exactly cancel out for all orders and the remaining finite value exactly matches with the left hand side (\ref{eren2l}) of the second Ehrenfest's relation (the mismatch between the last digit is only due to a rounding error). Finally to get the full picture we repeated the above analysis for those values of $\Omega$ ($0<\Omega<1$)for which the critical points lie within the physical domain and results clearly suggest that this phase transition is second order.


\section{Thermodynamic geometry}

Recently, analysis of black hole phase transitions from the point of view of thermodynamic state space (Ruppeiner) geometry has drawn some attention. In this approach Ruppeiner metric is defined by the Hessian of the entropy and this gives the pair correlation function \cite{Ruppeiner:1995zz, Ruppeiner:1983zz}. Also, the invariant length on the thermodynamic state space, defined by the Ruppeiner metric, when exponentiated, gives the probability distribution of fluctuations around the maximum entropy state \cite{Aman:2003ug}. Using the tools of Riemannian geometry, Ruppeiner curvature scalar ($R$) is defined and this is interpreted as the correlation volume multiplied by some proportionality constant \cite{Ruppeiner:2008kd}. Normally it is believed that the idea of correlation length has its root in the microscopic details of the system. But, remarkably, in the Ruppeiner geometry, a purely thermodynamic quantity ($R$) is claimed to serve the same purpose as the correlation length. This interpretation has been found quite successful in describing many second order phase transitions where this curvature scalar diverges at the critical point \cite{Shen:2005nu,Sahay:2010wi}. Now we examine whether such an interpretation holds for phase transition in Kerr AdS black hole.

The Ruppeiner metric is defined as \cite{Ruppeiner:1995zz}
\begin{equation}
dS^2=g_{ij}^{R}dX^idX^j
\label{rupp}
\end{equation} 
where, $g_{ij}=-\frac{\partial^2{S(X^k)}}{\partial X^i\partial X^j},~~{\textrm{and}}~~X^i\equiv X^i(M,N^a)$. Here $N^a$'s are all other extensive variables of the system. For Kerr AdS black hole $N^a=J$. In order to find $g_{ij}$ it is desirable to express $S$ in terms of $M$ and $J$. However from (\ref{M2}) we see that M is expressed as a function of $S$ and $J$ which is not invertible. In fact in this situation we can calculate the Weinhold metric which is defined in the following way \cite{Weinhold}, 
\begin{equation}
dS_{W}^2=g_{ij}^{W}dX^i dX^j
\label{wein}
\end{equation}
where $g_{ij}^{W}=\frac{\partial^2{M(X^k)}}{\partial X^i\partial X^j},~~{\textrm{and}}~~X^i\equiv X^i(S,J)$. It is well known that Ruppeiner metric and Weinhold metric are related by a conformal factor \cite{Mru, Ferrara}
\begin{equation}
dS_{R}^2=\frac{1}{T}dS_{W}^2
\label{confrel}
\end{equation}
where $T$ is the temperature of the system. In our example this would correspond to the Hawking temperature of the Kerr AdS black hole. Now using (\ref{M2}) we can easily calculate $dS_{W}^2$ and from that we can get $dS_{R}^2$ by using (\ref{confrel}) and (\ref{temp}). An explicit form of $dS_{R}^2$ is given by
\begin{equation}
dS_{R}^2=g_{SS}dS^2+2g_{SJ}dSdJ+g_{JJ}dJ^2
\label{metric}
\end{equation}
where
\begin{eqnarray}
g_{SS} &=&- \frac{1}{2}\left(-\frac{3}{S}-\frac{1}{\pi+S}-\frac{S(2\pi+3S)}{4J^2\pi^3+S^2(\pi+S)}+\frac{4S(\pi^2+6\pi S+6S^2)}{-4J^2\pi^4+S^2(\pi+S)(\pi+3S)}\right)\nonumber\\
g_{SJ} &=& g_{JS}=-4J\pi^3\left(-\frac{1}{4J^2\pi^3+S^2(\pi+S)}+\frac{2\pi}{4J^2\pi^4-S^2(\pi+S)(\pi+3S)}\right)\label{compon}\\
g_{JJ} &=& -\frac{4\pi^3S^3(\pi+S)^2}{(4J^2\pi^3+S^3+S^2\pi)(-4J^2\pi^4+S^2(\pi+S)(\pi+3S))}\nonumber
\end{eqnarray}
By definition Ruppeiner curvature is constructed exactly like the Riemannian curvature and for two dimensions the curvature scalar is given by \cite{Ruppeiner:1995zz},
\begin{equation}
R=-\frac{1}{\sqrt g}\left[\frac{\partial}{\partial S}\left(\frac{g_{SJ}}{\sqrt{g}g_{SS}}\frac{\partial g_{SS}}{\partial J}-\frac{1}{\sqrt g}\frac{\partial g_{JJ}}{\partial S}\right)+\frac{\partial}{\partial J}\left(\frac{2}{\sqrt g}\frac{\partial g_{SJ}}{\partial J}-\frac{1}{\sqrt g}\frac{\partial g_{SS}}{\partial J}-\frac{g_{SJ}}{{\sqrt g}g_{SS}}\frac{\partial g_{SS}}{\partial S}\right)\right]
\label{curv}
\end{equation}
Considering the metric (\ref{compon}) we now calculate $R$ as a function of $S$ and $J$. Finally using the relation (\ref{26}), one can find $R$ as a function of $S$ and $\Omega$. Since these expressions are quite lengthy we do not give those results here, instead we plot $R$ with $S$ for different values of $\Omega$, as shown in figure 3. 
\begin{figure}[h]
\centering
\includegraphics[angle=0,width=10cm,keepaspectratio]{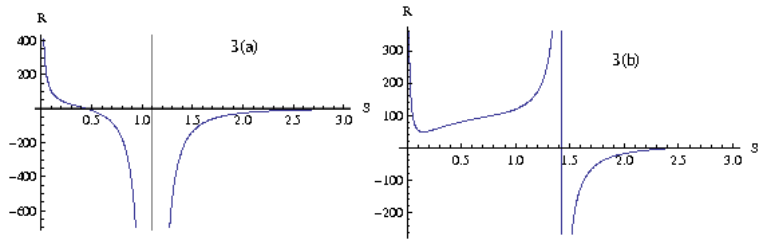}
\caption[]{{Thermodynamic scalar curvature ($R$) vs entropy ($S$) plot: $\Omega=0.3$ in 3(a) and $\Omega=1.5$ in 3(b).}}
\label{figure 5}
\end{figure}

Let us now explain the different plots one by one:\\
(i) In figure 3(a) we take $\Omega=0.3$ and this curve shows a divergence at $S_{\textrm{crit}}=1.0976$ which is exactly the case where $C_{\Omega}$ was discontinuous. Since we have already shown that the nature of this phase transition is second order it confirms the common belief that the divergence of $R$ means occurrence of second order phase transition.\\
(ii) In figure 3(b) we see that $R$ is also divergent at $S=1.4$ for $\Omega=1.5$. Note that this choice for $S$ and $\Omega$ is compatible with the consistency condition (\ref{cond}). But we do not find any discontinuity in $C_{\Omega}$ at this point and therefore there is no phase transition occurring here. This clearly shows that for this case it is incorrect to associate the divergence of $R$ with a phase transition.\\
Though $R$ is divergent in both figures 3(a) and 3(b), it is clear that the nature of divergences are completely different. While in figure 3(a) the divergent sector of the curve is symmetric with respect to the $R$ axis shifted at the singular point ($S_{\textrm{crit}}=1.0976$), in the other figure it is antisymmetric at the singular point. Thus from our analysis we conclude that a divergence of $R$ that is similar to figure 3(a) signals a second order phase transition in the Kerr AdS black hole.


\section{Conclusions}
Let us now summarise the findings of the present paper. Here we adopted the standard formalism, based on Clapeyron's and Ehrenfest's scheme used in conventional thermodynamic systems. According to Clapeyron's formalism a discontinuity in entropy with respect to the temperature is necessary for a first order phase transition. However the semiclassical entropy of the Kerr-AdS black hole was continuous and thus the possibility of a first order phase transition was absent. The discontinuity in the specific heat ($C_{\Omega}$) of the Kerr-AdS black hole suggested the possibility of a higher order phase transition and motivated us to check the validity of Ehrenfest's equations. For a true second order phase transition these equations must be satisfied. We discovered that for all allowed values of the angular velocity ($0<\Omega<1$), where the critical point lied in the physical domain, the Ehrenfest's equations were satisfied.

The differences in the application of Ehrenfest's scheme to black holes vis-a-vis conventional thermodynamical systems were highlighted. The infinite divergences appearing on the phase transition curves was discussed. We developed a method for checking the validity of Ehrenfest's relations infinitesimally close to the critical point. As a remarkable fact it was found that the infinite divergences of various physical quantities ($C_{\Omega},~\alpha,~k_T$) cancel each other eventually leading to a confirmation of both Ehrenfest's relations.

Another important part of our paper was an attempt to connect the state space geometry with the phase transition of the Kerr AdS black hole. We calculated the Ruppeiner curvature scalar ($R$) and investigated its behaviour at the critical point of a second order phase transition ($\Omega < 1$). It was found that $R$ always diverged at this critical point. This was compatible with a pattern, suggested by indivudual studies carried out for various thermodynamic systems (for a review see \cite{Ruppeiner:1995zz}), that the divergence of $R$ is a characteristic of a second order phase transition. However, a divergence in $R$ was also noted for $\Omega\geq 1$ where no phase transition occurs. This indicated a deviation from the aforestated pattern where a divergence in $R$ signalled a phase transition. But it must be stressed that the nature of the divergence of $R$ in the latter ($\Omega\geq 1$) case was found to be completely different from the previous ($\Omega < 1$) case. This is illustrated by looking at figure 3 (fig. 3(a) corresponds to $\Omega<1$ while fig. 3(b) corresponds to $\Omega\ge 1$). Thus our study revealed that only a particular type of singularity (symmetric type) in $R$ can correctly locate the phase transition point for the Kerr-AdS black hole.\\

{\it Acknowledgement}: One of the authors (S.K.M) thanks the Council of Scientific and Industrial Research (C.S.I.R), Government of India, for financial support.


\end{document}